\documentclass[letterpaper,prl,aps,twocolumn,showpacs,superscriptaddress]{revtex4-1}
\usepackage{bm,color}
\usepackage{mathptmx}
\usepackage[T1]{fontenc}
\usepackage[utf8]{inputenc}
\usepackage{graphicx} 
\usepackage{amsmath} 
\usepackage{amsfonts} 
\usepackage{braket}

	\newcommand{\bbm}{\begin{pmatrix}}
	\newcommand{\ebm}{\end{pmatrix}}

	\usepackage[normalem]{ulem} 
	\definecolor{mgreen}{RGB}{1,123,0}

\begin{document}

\title{High-precision multiband spectroscopy of ultracold fermions in a non-separable optical lattice}

\author{Nick Fläschner}
\affiliation{Institut für Laserphysik, Universität Hamburg, 22761 Hamburg, Germany}
\affiliation{The Hamburg Centre for Ultrafast Imaging, 22761 Hamburg, Germany}
\author{Matthias Tarnowski}
\affiliation{Institut für Laserphysik, Universität Hamburg, 22761 Hamburg, Germany}
\affiliation{The Hamburg Centre for Ultrafast Imaging, 22761 Hamburg, Germany}
\author{Benno S. Rem}
\affiliation{Institut für Laserphysik, Universität Hamburg, 22761 Hamburg, Germany}
\affiliation{The Hamburg Centre for Ultrafast Imaging, 22761 Hamburg, Germany}
\author{Dominik Vogel}
\affiliation{Institut für Laserphysik, Universität Hamburg, 22761 Hamburg, Germany}
\author{Klaus Sengstock}
\email{klaus.sengstock@physnet.uni-hamburg.de}
\affiliation{Institut für Laserphysik, Universität Hamburg, 22761 Hamburg, Germany}
\affiliation{The Hamburg Centre for Ultrafast Imaging, 22761 Hamburg, Germany}
\affiliation{Zentrum für Optische Quantentechnologien, Universität Hamburg, 22761 Hamburg, Germany}
\author{Christof Weitenberg}
\affiliation{Institut für Laserphysik, Universität Hamburg, 22761 Hamburg, Germany}
\affiliation{The Hamburg Centre for Ultrafast Imaging, 22761 Hamburg, Germany}

\date{\today}


\begin{abstract}
Spectroscopic tools are fundamental for the understanding of complex quantum systems. Here we demonstrate high-precision multi-band spectroscopy in a graphene-like lattice using ultracold fermionic atoms. From the measured band structure, we characterize the underlying lattice potential with a relative error of $1.2\cdot 10^{-3}$. Such a precise characterization of complex lattice potentials is an important step towards precision measurements of quantum many-body systems. Furthermore, we explain the excitation strengths into the different bands with a model and experimentally study their dependency on the symmetry of the perturbation operator. This insight suggests the excitation strengths as an suitable observable for interaction effects on the eigenstates.
\end{abstract}

\maketitle

Cold atoms in optical lattices are a versatile platform to study general lattice physics phenomena in a well-controlled environment and with tunable interactions \cite{Lewenstein2007, Bloch2008}. Recently, progress has been made in the direction of more complex lattice geometries and tunable lattices that bring the field closer to the complexity of solid-state physics \cite{Windpassinger2013}. Examples include artificial graphene \cite{Soltan-Panahi2011, Tarruell2012}, Kagome lattices \cite{Jo2012}, quadratic superlattices \cite{Aidelsburger2011, Miyake2013}, Lieb lattices \cite{Taie2015}, spin-orbit coupled lattices \cite{Wu2016}, and different highly-tunable lattices \cite{Soltan-Panahi2011, Tarruell2012, Weinberg2016, Flaschner2016, Salger2007}. With increasing complexity and tunability of the optical lattices, it becomes more and more relevant to have a precise measurement of the single-particle band structures to characterize the underlying lattice potential. This knowledge is a necessary prerequisite for bringing the study of strongly-interacting many-body phases in optical lattices to the same high-precision regime that is now reached in continuous systems \cite{Ku2012, Desbuquois2014}. Moreover, the precise knowlegde of the band structure is important as starting point for Floquet engineering of optical lattices \cite{Lignier2007, Struck2011, Eckardt2017}, e.g. for creating topological bands \cite{Jotzu2014, Aidelsburger2015, Tarnowski2017b}. In this context, the characterization of the higher bands is crucial for understanding and avoiding heating resonances into these bands \cite{Weinberg2015, Reitter2017}.

A variety of spectroscopic tools have been applied to quantum gases in optical lattices including Bragg spectroscopy \cite{Ernst2010,Fabbri2012, Ha2015}, spin-injection spectroscopy \cite{Cheuk2012, Weinberg2016, Huang2017}, Fourier transform spectroscopy \cite{Valdes-Curiel2017}, Stückelberg interferometry \cite{Kling2010, Li2016}, and amplitude modulation spectroscopy \cite{Stoeferle2004, Heinze2011}. Here we demonstrate a fully momentum-resolved amplitude modulation spectroscopy of a non-separable two-dimensional lattice. We use a honeycomb optical lattice, which realizes artificial graphene. We optimize the experimental protocol to obtain excellent data quality for the full two-dimensional non-separable band structure, which allows the determination of the underlying lattice potential to unprecedented precision.

	\begin{figure}[h]
		\includegraphics[width=\linewidth]{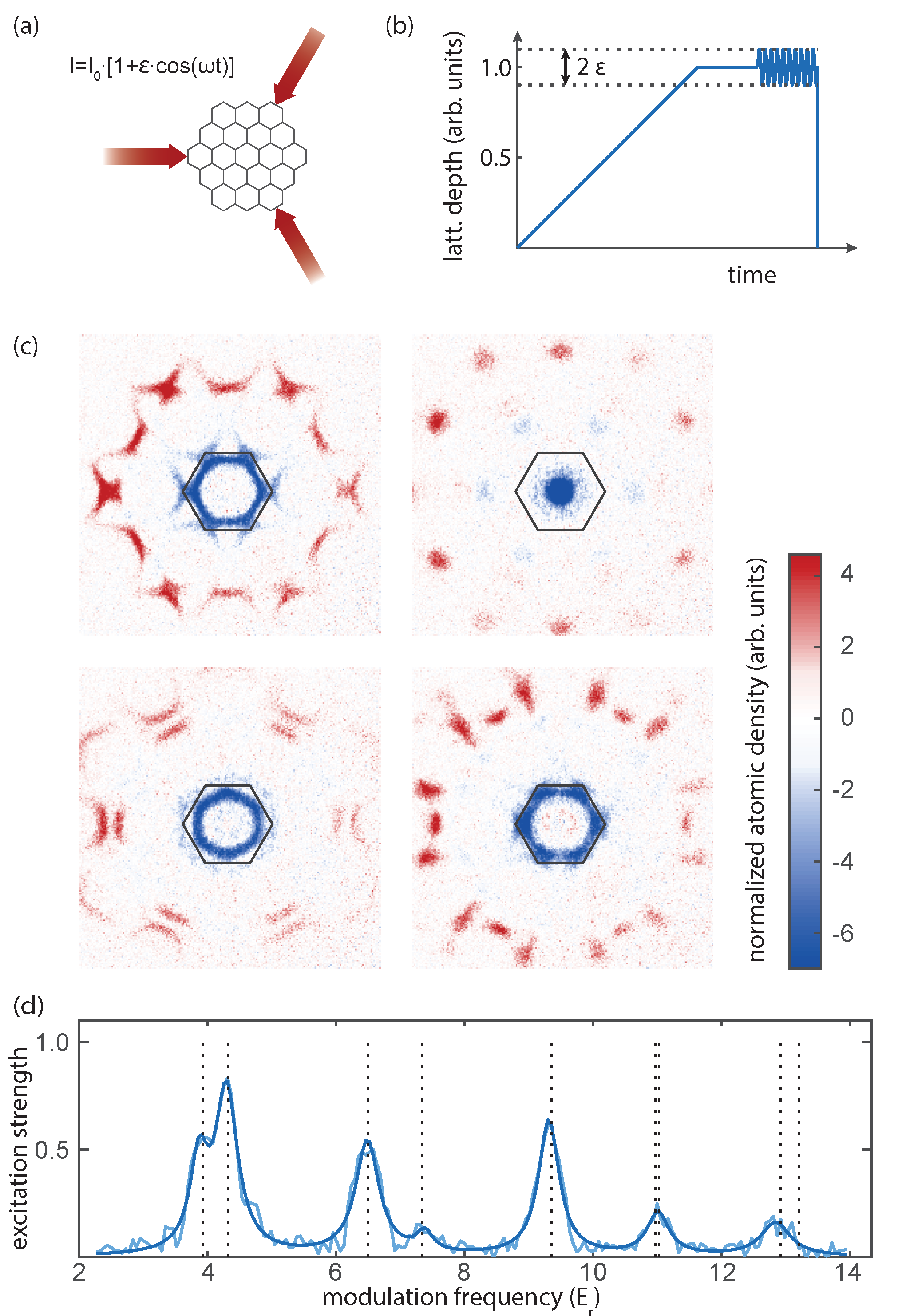}
		\caption{Amplitude modulation spectroscopy in a honeycomb lattice. (a) Setup of the honeycomb lattice formed by three interfering laser beams (red arrows) (b) Sketch of the lattice depth ramp in the experimental sequence inclucing the adiabatic loading and the modulation with amplitude $\epsilon=0.1$. (c) Four examples of time-of-flight pictures after resonant modulation ($\hbar\omega=5.84, 8.90, 9.75, 10.43\,E_r$). A picture without excitations is substracted such that the missing atoms in the first Brillouin zone (grey hexagon) are clearly visible. These images are obtained for symmetric modulation. (d) Excitation spectrum for one quasimomentum (i.e. one pixel of the images) as function of the modulation frequency. The data (light blue) is well described by a fit (dark blue) consisting of a sum of Lorentzians. The peaks occur at the positions expected from the band structure (dashed black line, band structure from fit to all data). This spectrum is a weighted average over symmetric and asymmetric modulation in order to access almost all bands. All data in this article is averaged over four different modulation times 500, 510, 520, 530 $\mu$s. See \cite{SupMat} for details.}\label{fig:1_System}
		\end{figure} 

The hexagonal optical lattice is formed by three interfering laser beams ($\lambda=1064\,$nm) intersecting under 120$^{\mathrm{o}}$. For the choice of balanced intensities and for in-plane polarization, a graphene-like honeycomb lattice is realized \cite{Soltan-Panahi2011} (Fig.\,1). The experiments start with a cold sample of spin-polarized fermionic $^{40}$K atoms in a crossed optical dipole trap (recoil energy $E_r=h\cdot 4.41\,$kHz). A completely filled lowest band is prepared by adiabatically ramping up the lattice depth. We then create excitations into higher bands by sinusoidal amplitude modulation of the lattice depth. This perturbation conserves quasimomentum. Because the higher bands have a different curvature than the lowest band, the excitation frequencies are quasimomentum dependent. Therefore each modulation frequency will only create excitations at certain quasimomenta, making the method momentum-resolved. After the modulation, the lattice potential is abruptly switched off and after a time-of-flight expansion, the momentum distribution is obtained. The excitations show as reduced densities in the first Brillouin zone (Fig\,1c). 

In order to reach the required high precision and to avoid distortion, we do not apply adiabatic band mapping and thus lose the information into which band we excite. None-the-less the excitations can be clearly identified as holes in the first Brillouin zone, because the Bloch coefficients for the momenta in the first Brillouin zone of higher bands are smaller than those of the lowest band. The excitations also appear as increased densities at higher real momenta, but we restrict the analysis to the holes in the first Brillouin zone to minimize the effect of wavepacket dynamics during the excitation pulse induced by the external trap \cite{Sherson2012, Heinze2013}. The trap-induced dynamics also leads to a closing of the holes in the first Brillouin zone, but this effect does not change the hole position and the relative hole depth still reaches values up to 75\%, because the excitation rate is larger than the typical hole closing rate on the order of 200 Hz, which is expected for our trapping frequency of $70\,\mathrm{Hz}$ \cite{Heinze2013}. These details of the measurement protocol and of the data evaluation are crucial for avoiding systematic errors and for reaching the high precision.

In order to obtain the spectra, we repeat the modulation for different frequencies and monitor the excitation strength for each quasimomentum (Fig.\,1d). In contrast to previous work, where we used the resulting coherences between the lowest band and the excited bands to obtain information about the eigenstates of the lattice \cite{Tarnowski2017}, we here wash out these coherences by averaging the signal over several modulation times.

	\begin{figure*}
		\includegraphics[width=\textwidth]{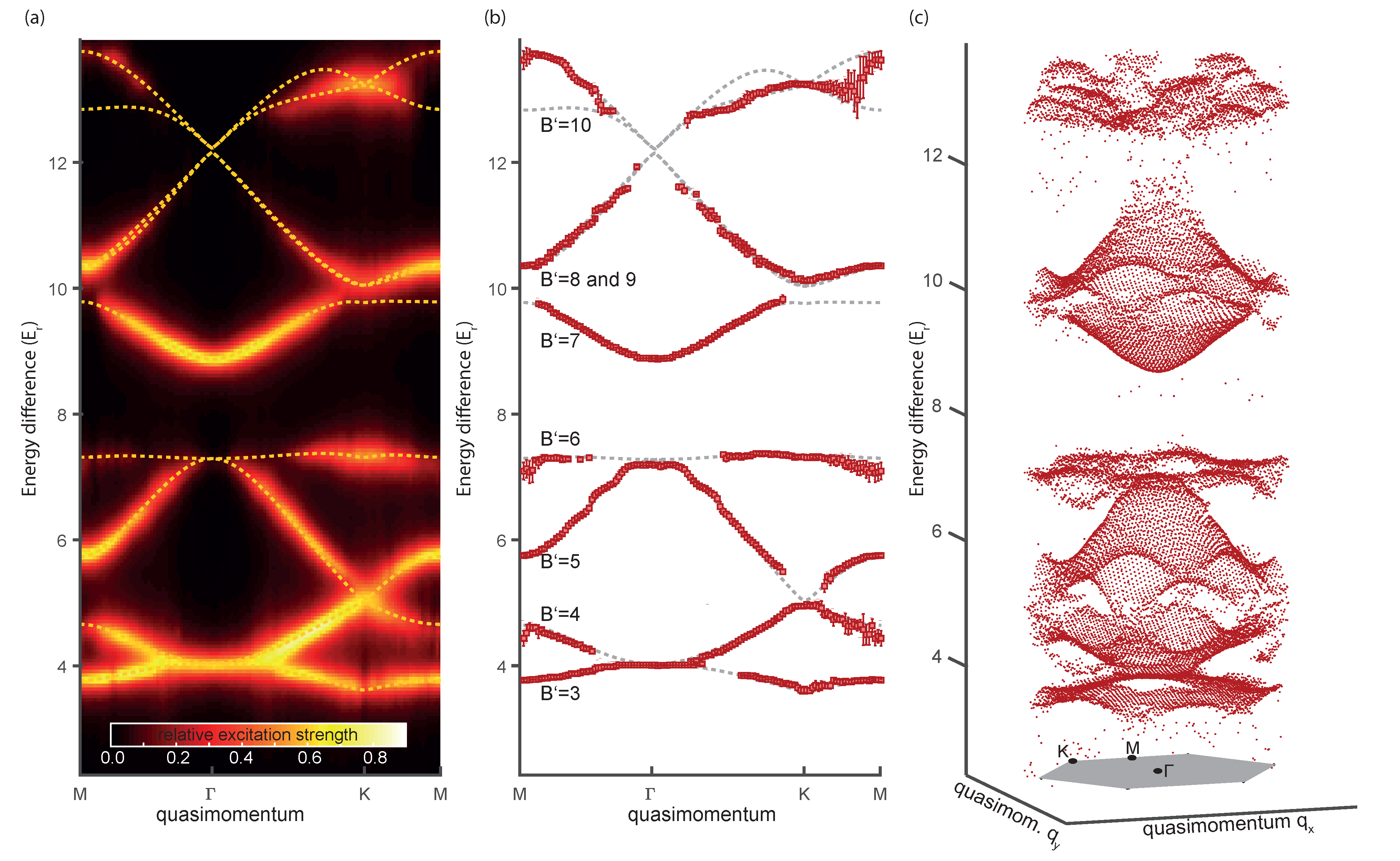}
		\caption{Measured band structure of the honeycomb lattice. (a) Excitation spectrum across a high symmetry path in the first Brillouin zone (averaged over the six equivalent paths). The data is filtered via a fit as described in Fig.\,1. Yellow dashed lines indicate the calculated band structure for the fitted lattice depth of $\tilde{V}=8.57\,E_r$ which is valid for all data in this article. (b) Extracted maxima of the spectrum in a. The error bar denotes the standard deviation from averaging over the six equivalent paths. The missing data points at some momenta (e.g. at the crossing between bands 8, 9 and 10) are due to coupling strength via amplitude modulation, which vanishes here (see Fig.\,3). (c) Extracted maxima of the complete two-dimensional band structure. The grey hexagon marks the first Brillouin zone and the high symmetry points used in a and b are indicated as black dots.}\label{fig:2_ExcitationSpectra}
	\end{figure*}

The resulting spectra along a high-symmetry path through the Brillouin zone are shown in Fig.\,2. From the maxima of the excitation strength, the full two-dimensional band structure can be obtained (Fig. 2b,c). The spectrum shows the four p-bands (with band index B'=3...6) and five of the d-bands (B'=7..11). The p-bands are well separated from the rest of the spectrum and show the well-known structure of two outer narrow bands and two inner bands, which linearly touch in Dirac points \cite{Wu2008}. Note that the spectrum shows the energy difference to the lowest band, which is itself curved (the lowest band has a width of $0.4\,E_r$).

The observed full width at half maximum of the excitations is on the order of $0.5 E_r\approx h \cdot 2\,$kHz in energy. This width is limited by the 2\,kHz Fourier width of our rectangular spectroscopy pulse of duration $T=0.5\,\mathrm{ms}$. 
Furthermore, the width is affected by the inhomogeneity of the lattice potential, which results from the finite waist of the lattice beams ($1/e^2$ radius of 160\,$\mu$m). In a local density approximation, one can define a local lattice depth, which varies across the system by about 3\% between the center and the edges at a radius of about 20\,$\mu$m. We state in this article the mean lattice depth $\tilde{V}$, which is experienced by most atoms and which can be determined to a much greater precision than this variation.  

The large amount and high quality of the data allows for a very precise fit of a calculated band structure and thereby for a very precise determination of the underlying lattice potential characterized by $\tilde{V}$. We determine $\tilde{V}=(8.57 \pm 0.01)\,E_r$ to a relative standard error of $1.2\cdot10^{-3}$, which is an improvement by a factor of 15 compared to previous work \cite{Heinze2011}.

	\begin{figure*}
		\includegraphics[width=\textwidth]{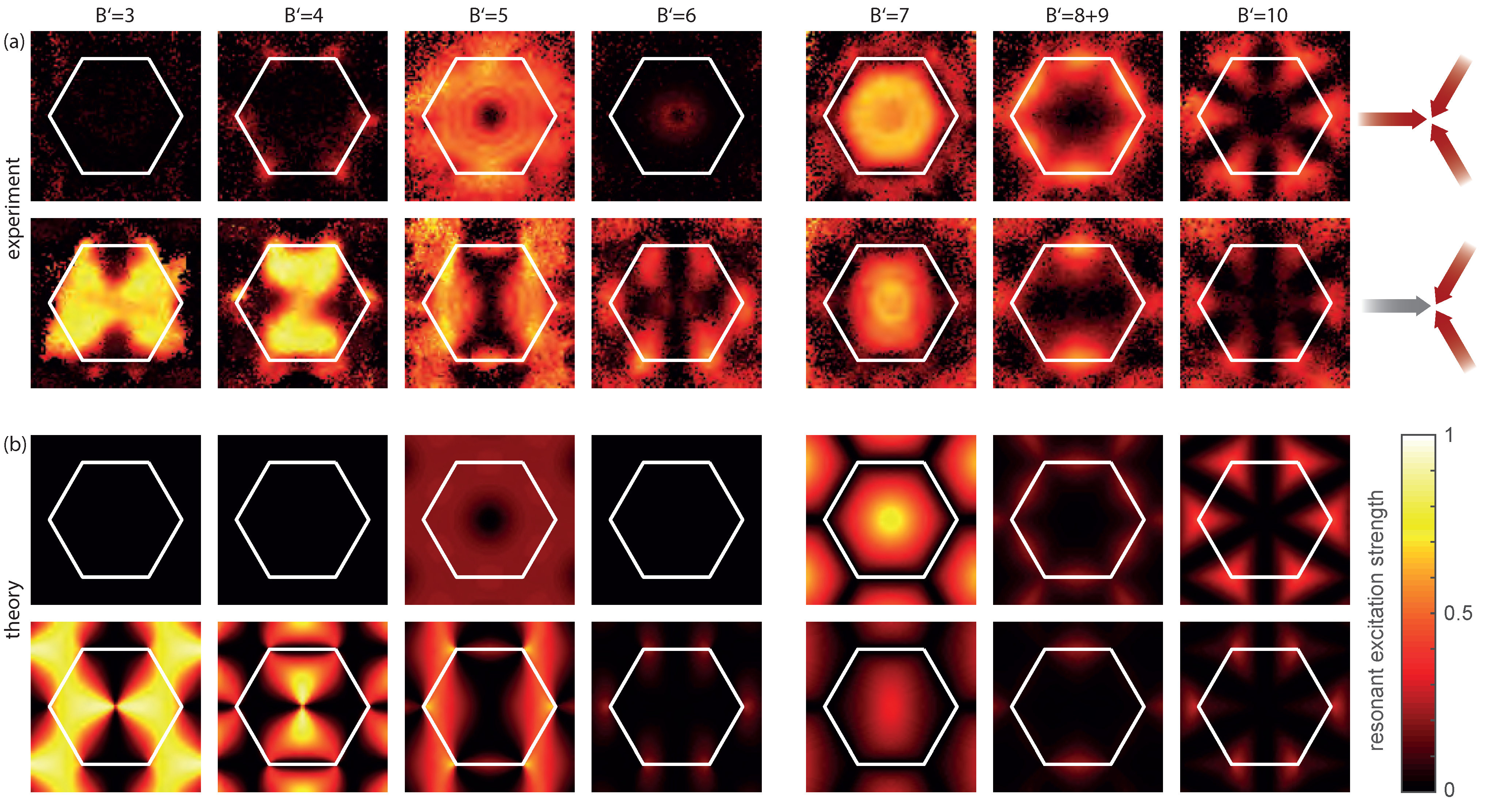}
		\caption{Resonant excitation strengths for different symmetries of the perturbation operator. (a) Measured resonant excitation strengths from the lowest band into different higher bands across the full Brillouin zone (white hexagon) for symmetric modulation (top row) and asymmetric modulation (bottom row, one lattice beam is not modulated, indicated in grey in the sketch to the right). The excitation strength is evaluated at the excitation frequency expected from the band structure. (b) Calculated resonant excitation strengths using the model discussed in the text and the Bloch coefficients from the fitted band structure.
	}\label{fig:3_ExcitationStrength}
	\end{figure*}

In addition to the band structure, we also have access to the resonant excitation strengths from the lowest band into the different higher bands. In Fig.\,3 we plot the measured resonant excitation strengths for each band at the excitation frequencies expected for the lattice depth determined previously. We show this separately for symmetric modulation (all three lattice beams are modulated with equal amplitude) and for asymmetric modulation (only two of the lattices beams are modulated) and find very different patterns. While in the symmetric case, some bands are not accessible, this changes drastically for asymmetric modulation. This suggests asymmetric modulation as a straight forward strategy to reach these bands. 

These observations can be understood with a clear physical picture. Optical lattices arise from Bragg processes between two lattice beams, which couple plane wave states differing by a reciprocal lattice vector. The amplitude modulation makes additional processes resonant, which also change the band index. The matrix element for these processes is then given by the overlap of the initial Bloch state shifted by one reciprocal lattice vector with the final state in the higher band \cite{SupMat}. The results of this model using the Bloch coefficients from the band structure fit are shown in Fig. 3 and match very well with the data.  

In our non-separable lattice formed by three lattices beams, three such processes occur (one for each pair of lattice beams, i.e. for a shift along each of the three reciprocal lattice vectors) and their strengths can be independently tuned via the symmetry of the perturbation operator. The three processes can interfere constructively or destructively. The zero excitation strength for symmetric modulation is in fact mostly due to destructive interference of these three processes, as we explicitly demonstrate by re-establishing the coupling into these bands using asymmetric modulation. This effect is particularly striking for three of the four p-bands, where the excitation is strictly zero for symmetric modulation, but becomes very large for asymmetric modulation. These observations are a nice example of quantum interference between different excitation paths that arises in many areas of quantum physics, e.g. in atomic dark states \cite{Cohen-Tannoudji1991}. The independent control of the paths is naturally given in our non-separable lattice.

In a few cases, all three processes vanish individually and the excitation strength remains zero for all perturbation symmetries. This is e.g. true for B'=5 and q=0 and for the complete upper s-band (B'=2 not shown). In these cases, the two sub-processes of shifting by plus or minus the given reciprocal lattice vector interfere destructively, which cannot be suppressed by changing the symmetry of the perturbation operator. The corresponding excited states can then be addressed via other perturbations such as lattice shaking. 

In conclusion, we have presented the first momentum-resolved multiband spectroscopy in a non-separable two-dimensional lattice. We compared both the excitation frequencies and the resonant excitation strengths with a band structure calculation and find excellent agreement. The data quality allows determining the underlying lattice potential with a relative error of $1.2\cdot 10^{-3}$, which is an important step towards precision many-body quantum simulation in optical lattices.

Higher band physics has recently attracted much attention due to the appearance of new phases such as chiral superfluidity \cite{Wirth2011,Kock2015}. Moreover, the higher bands could be used as probe bands to characterize strongly-correlated phases in the ground band. In analogy to rf spectroscopy into a third, non-interacting internal state, one could perform precision spectroscopy into higher bands, where the atoms interact much less with the atoms in the ground band due to the smaller density overlap of the Wannier functions of different symmetry \cite{Bakr2011}. This would be particularly useful for detecting the dispersion of interaction-induced quasi-particles.

We further demonstrated that each higher band is accessible via amplitude modulation by adjusting the symmetry of the perturbation operator, which can be achieved by modulating a subset of the lattice beams. We explain these findings with a model based on the overlap of Bloch states shifted by one reciprocal lattice vector. The resonant excitation strength can thus be directly calculated from the eigenstates and might therefore be a good observable for identifying interaction effects on the eigenstates of the lowest band, similar to the effects on the energies studied before \cite{Heinze2011}. One could use these insights into the coupling strength to design Floquet protocols that selectively couple to a single band and therefore avoid heating into the other bands \cite{Weinberg2015, Reitter2017}. Using the excitation strengths for learning about the eigenstates has recently also been proposed for lattice shaking spectroscopy, where the topology of the lowest band can be inferred \cite{Tran2017}.

\begin{acknowledgments}
We acknowledge financial support from the Deutsche Forschungsgemeinschaft via the Research Unit FOR 2414 and the excellence cluster “The Hamburg Centre for Ultrafast Imaging - Structure, Dynamics and Control of Matter at the Atomic Scale”. BSR acknowledges financial support from the European Commission (Marie Curie Fellowship).
\end{acknowledgments}


%


\clearpage
\appendix

\section{SUPPLEMENTAL MATERIAL}
\section{Calculation of the resonant excitation strength}
The non-separable potential of the optical honeycomb lattice can be written as the sum of three non-orthogonal one-dimensional lattices
\begin{align}
\label{latticedef}
V(\mathbf{r})&=V_{12}(\mathbf{r})+V_{23}(\mathbf{r})+V_{13}(\mathbf{r}) \nonumber \\&= V_{12}\cos\left({\mathbf{b_1}\cdot\mathbf{r}}\right)+V_{23}\cos\left({\mathbf{b_2}\cdot\mathbf{r}}\right)+V_{13}\cos\left({\mathbf{b_3}\cdot\mathbf{r}}\right)
\end{align}
with the reciprocal lattice vectors $\mathbf{b_1}=\sqrt{3}k_L\left(1/2,\sqrt{3}/2\right)$ and $\mathbf{b_2}=\sqrt{3}k_L\left(-1,0\right)$ and $\mathbf{b_3}=\mathbf{b_1}+\mathbf{b_2}$, the lattice beam wave number $k_L=2\pi/1064\,$nm and the one-dimensional lattice depths $V_{ij}$ being proportional to $V_{ij}\propto\sqrt{I_iI_j}$ with the intensities $I_i$ of the lattice laser beams. The eigenstates of the  lattice Hamiltonian $\hat{\mathbf{H}}=\hat{\mathbf{p}}^2/2m+V(\mathbf{r})$ are the spatially periodic Bloch states $\Psi_{\mathbf{q},B}(\mathbf{r})$,
\begin{equation}
\label{blochtheorem}
\Psi_{\mathbf{q},B}(\mathbf{r})=\exp\left(-i\mathbf{q}\cdot\mathbf{r}\right) \sum_{\mathbf{K}}c_{\mathbf{K}}^{\mathbf{q},B}\exp{\left(-i\mathbf{K}\cdot\mathbf{r}\right)}
\end{equation}
which are labeled by the quasimomentum $\mathbf{q}$ and the band index B. The $c_{\mathbf{K}}^{\mathbf{q},B}$ are the Bloch coefficients, which can be obtained from a band structure calculation. The summation is performed over all reciprocal lattice vectors $\mathbf{K}=n_1\mathbf{b_1}+n_2\mathbf{b_2}$ with $n_1,n_2 \in \mathbb{Z}$. Amplitude modulation spectroscopy is performed by periodically modulating the intensity of the lattice laser beams, 
\begin{equation}
I_i(t)=I_{i,0}+\epsilon_i I_{i,0}\sin{\left(\omega t\right)}
\end{equation} 
where $\epsilon_i$ is the modulation strength and $\omega$ the spectroscopy frequency. This leads to a time-dependent lattice depth of the respective one-dimensional lattices, to first order
\begin{equation}
V_{ij}(\mathbf{r},t)\approx V_{ij}(\mathbf{r})+\frac{1}{2}\left(\epsilon_i+\epsilon_j\right)\sin{\left(\omega t\right)}V_{ij}(\mathbf{r})
\end{equation}
such that the full two-dimensional time-dependent lattice potential can be written as
\begin{widetext}
\begin{align}
V(\mathbf{r},t)&=V(\mathbf{r})+\frac{1}{2}\sin{\left(\omega t\right)}\left[ \left(\epsilon_1+\epsilon_2\right)V_{12}(\mathbf{r})+ \left(\epsilon_2+\epsilon_3\right)V_{23}(\mathbf{r})+ \left(\epsilon_1+\epsilon_3\right)V_{13}(\mathbf{r})\right] = V(\mathbf{r})+\sin{\left(\omega t\right)} V'(\mathbf{r}),
\label{timedep}
\end{align}
\end{widetext}
where we have defined $V'(\mathbf{r})$ as the time-independent perturbation operator. 

In order to calculate the resonant excitation strength as plotted in Fig.\,3 of the main text, we make use of time-dependent perturbation theory, assume an infinite perturbation time and use Fermi's golden rule. Then, the probability $P(\mathbf{q},\mathbf{q'},B=1,B')$ that an atom in the initial state with quasimomentum $\mathbf{q}$ and band index $B=1$ is excited to quasimomentum $\mathbf{q'}$ and band $B'$ is to first order given as 
\begin{equation}
\label{fermisrule}
P(\mathbf{q},\mathbf{q'},B=1,B')=\delta_{E,\hbar \omega} \left| \int d^3 r \Psi_{\mathbf{q},1}(\mathbf{r}) V'(\mathbf{r}) \Psi^{\ast}_{\mathbf{q'},B'}(\mathbf{r})\right|^2
\end{equation}
where $E$ is the energy difference between the two states. Plugging equations (\ref{latticedef}) and (\ref{timedep}) into (\ref{fermisrule}) on resonance ($E=\hbar \omega$) yields 

\begin{widetext}
\begin{eqnarray}
\label{overlapresult}
P(\mathbf{q},\mathbf{q'},B=1,B')=\delta_{\mathbf{q},\mathbf{q'}} \bigg\vert &\frac{V_{12}}{4}&\left(\epsilon_1+\epsilon_2\right)\sum_{\mathbf{K}} c_{\mathbf{K}}^{\ast,\mathbf{q'},B'}\left(c_{\mathbf{K}+\mathbf{b_1}}^{\mathbf{q},1}+ c_{\mathbf{K}-\mathbf{b_1}}^{\mathbf{q},1}\right) \\
+ &\frac{V_{23}}{4}&\left(\epsilon_2+\epsilon_3\right)\sum_{\mathbf{K}} c_{\mathbf{K}}^{\ast,\mathbf{q'},B'}\left(c_{\mathbf{K}+\mathbf{b_2}}^{\mathbf{q},1}+ c_{\mathbf{K}-\mathbf{b_2}}^{\mathbf{q},1}\right) \\
+ &\frac{V_{13}}{4}&\left(\epsilon_1+\epsilon_3\right)\sum_{\mathbf{K}} c_{\mathbf{K}}^{\ast,\mathbf{q'},B'}\left(c_{\mathbf{K}+\mathbf{b_3}}^{\mathbf{q},1}+ c_{\mathbf{K}-\mathbf{b_3}}^{\mathbf{q},1}\right) \bigg\vert ^2.
\end{eqnarray}
\end{widetext}
This expression shows that amplitude modulation is quasimomentum-preserving and can be understood as simply shifting the initial state in momentum space by a reciprocal lattice vector $\mathbf{b_1},\mathbf{b_2},\mathbf{b_3}$, as stated in the main text. The resonant excitation strength can thus be readily calculated from the Bloch coefficients $c_{\mathbf{K}}^{\mathbf{q},B}$. Furthermore, the result shows that the excitation process can be thought of as a quantum interference between the three processes stemming from the three one-dimensional lattices. Each of those three processes in turn consists of two sub-processes, namely the absorption and emission of one reciprocal lattice vector of the respective one-dimensional lattice. As explained in the main text, the absence of excitation can be due to two reasons: first, the two sub-processes of each one-dimensional lattice can interfere destructively, as is the case for, e.g., the second band ($B'=2$). Then, the states cannot be coupled by amplitude modulation in linear response. Second, the three processes originating from the three one-dimensional lattices can destructively interfere, as is the case for many excited states for, e.g., symmetric modulation ($\epsilon_1=\epsilon_2=\epsilon_3$). Since we can control the modulation strength $\epsilon_i$ of each beam individually, we can instead perform the modulation asymmetrically by, e.g., setting one $\epsilon_i=0$ which allows exciting into these states using amplitude modulation. Similar results are obtained by modulating only one lattice beam.

\section{Data analysis}
As stated in the main text, we perform amplitude modulation spectroscopy for four different spectroscopy times ($t=500, 510, 520, 530 \,\mathrm{\mu} s$) and for four different symmetries of the perturbation operator ($\epsilon_1=\epsilon_2=\epsilon_3=0.1$ and the three combinations with one $\epsilon_i=0.0$ and the other two $\epsilon_j=0.1$). The data is averaged over four different spectroscopy times, in order to wash out interference between the bands. The separation of 10 us between consecutive modulation times is chosen to be smaller than the shortest modulation period of 16 us (at the highest modulation frequency of 60 kHz). This choice is motivated by the need to avoid aliasing effects.
 
We first normalize all pictures to the same atom number to account for atom number fluctuations, we then average over four runs per parameter and then over the four spectroscopy times. Resulting data is shown in Figs.\,1c and 3 in the main text, where data for symmetric (Fig.\,1c and 3) and asymmetric modulation (Fig.\,3) is shown. For the data shown in Figs.\,1d and 2, we additionally average over all four modulation symmetries. In this average, we improve the signal to noise ratio by using a weighted average with the weights being the respective excitation strengths. We then fit a sum of Lorentzians to the data in order to filter the data and to extract resonance positions and widths. The fit to the (unfiltered) data is initialized with the positions and widths of the eight most prominent peaks found in Fourier-filtered data by the "findpeaks" routine in MATLAB. Fitted resonances whose full width at half maximum is below the Fourier limit or larger than one recoil energy are removed. Fig.\,1d in the main text shows a comparison of the data and the resulting fit for an exemplary momentum. The fitted resonance positions for all momenta are shown in Figs.\, 2b and c in the main text. From the fitted resonance positions, we estimate the lattice depth $V=V_{12}=V_{23}=V_{13}$ to a relative uncertainty of $0.1\%$, as explained in the following. 

\begin{figure}[htbp]
		\includegraphics[width=0.5\linewidth]{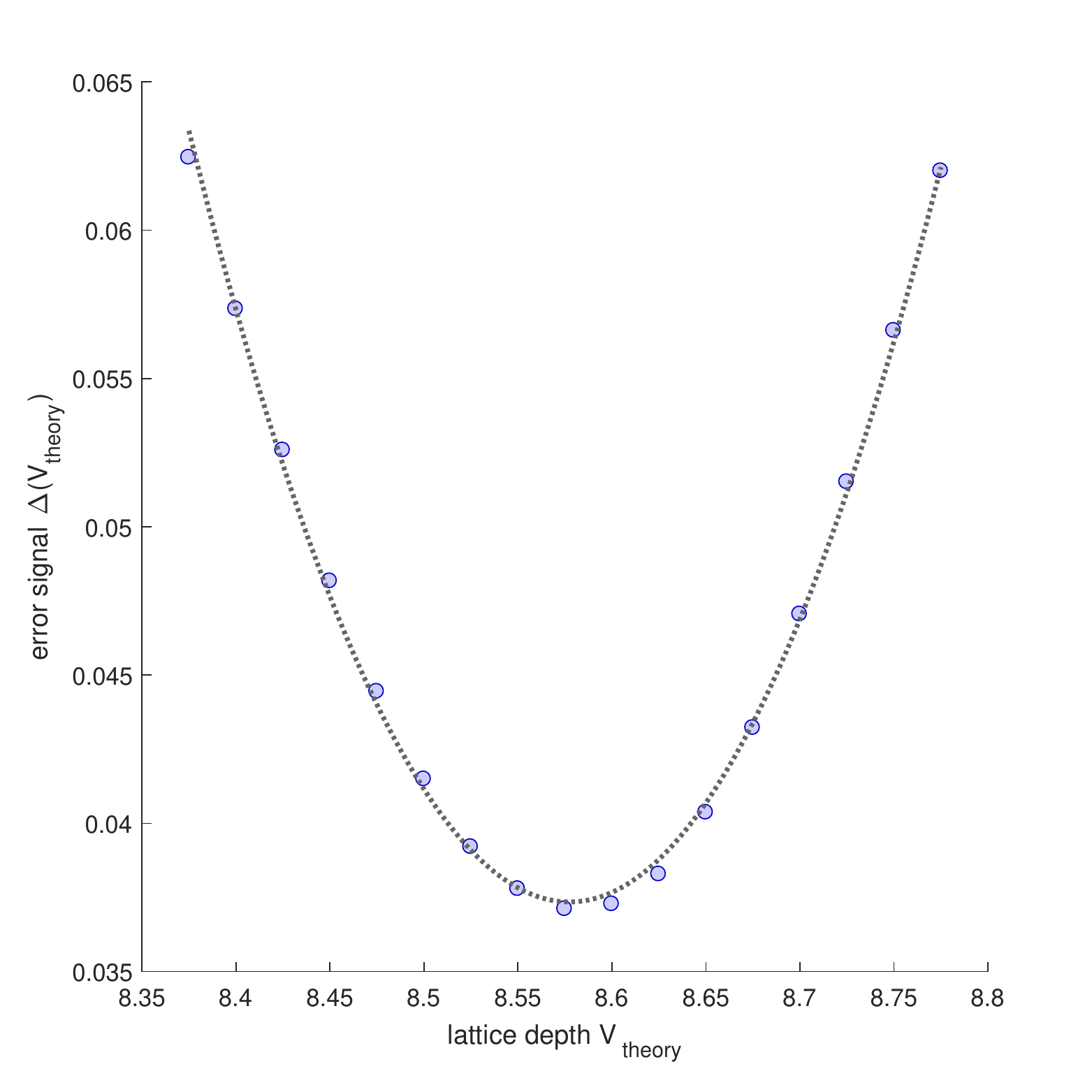}
		\caption{Determination of the experimental lattice depth. The error signal $\Delta(V_{\mathrm{theory}})$ is plotted (blue dots) as a function of the lattice depth. A parabolic fit (dashed line) yields the experimental lattice depth $V=(8.57\pm 0.01)\,E_{\mathrm{recoil}}$  }\label{fig:SM}
		\end{figure}

The error signal $\Delta(V_{\mathrm{theory}})$ which is (as a function of the lattice depth) minimized in order to obtain the experimental lattice depth is the mean deviation of all experimentally found resonances $E_{\mathrm{exp}}$ to an exact band structure calculation $E^{B,\mathbf{q}}_{\mathrm{theory}}$,
\begin{equation}
\Delta(V_{\mathrm{theory}})=\sum^{N}_{n=1} \bigg\vert\frac{E^{B,\mathbf{q}}_{\mathrm{theory}}(n)-E_{\mathrm{exp}}(n)}{N}\bigg\vert,
\end{equation}
where $N$ is the number of experimentally found resonances across all bands and momenta. Note that we omitted the weighting of the mean square deviation by the uncertainty of the experimental points, i.e. the width of the resonance peaks, because it is practically constant over all points. For each experimentally found resonance, we determine the band index $B$ by identifying the band which is energetically closest and deduce the quasimomentum $\mathbf{q}$ from the position and size of the first Brillouin zone. Prior to this procedure, obvious outliers $\big\vert E^{B,\mathbf{q}}_{\mathrm{theory}}(n)-E_{\mathrm{exp}}(n) \big\vert \geq 4 \Delta(V_{\mathrm{theory}})$ are removed, discarding approximately $4\%$ of the experimentally found resonances and leading to typical values of $N\approx 15300$. A further fit of a parabola to the error signal then yields the experimental lattice depth $V$ (Fig.\, \ref{fig:SM}). 
The determined experimental lattice depth is $V=8.57\pm 0.01\, E_{\mathrm{recoil}}$, where the uncertainty $\delta V$ is given by the curvature $a$ of the parabola and the number of data points $N$, $\delta V=1/\sqrt{aN}=0.01\,E_{\mathrm{recoil}}$, corresponding to a relative uncertainty of $0.1\%$.

The data set comprises about 8500 images, which corresponds to about two days of continuous measurements. This arises from 210 excitation frequencies, 4 different symmetries of the perturbation, 4 different modulation times and up to four averages under identical conditions. The error on the lattice depth is given by the number of extracted maxima, i.e. the number of quasimomenta times the number of measured higher bands. Therefore a high precision might also be reached with a coarser frequency resolution, which suggests our method as a practical tool for precise lattice calibration.

\end{document}